\documentclass[iop]{emulateapj}

\usepackage{amsmath}

\makeatletter
\newcommand\tsb[1]{\@textsubscript{\selectfont#1}}
\def\@textsubscript#1{{\m@th\ensuremath{_{\mbox{\fontsize\sf@size\z@#1}}}}}
\newcommand\tsp[1]{\@textsuperscript{\selectfont#1}}
\def\@textsuperscript#1{{\m@th\ensuremath{^{\mbox{\fontsize\sf@size\z@#1}}}}}

\usepackage{graphicx,color}
\providecommand{\edit}[1]{{\color{black}{#1}}}

\usepackage{natbib}
\usepackage[colorlinks, citecolor=blue,linkcolor=blue,urlcolor=blue]{hyperref}
\usepackage[all]{hypcap}
\usepackage{etoolbox}

\makeatletter

\patchcmd{\NAT@citex}
  {\@citea\NAT@hyper@{%
     \NAT@nmfmt{\NAT@nm}%
     \hyper@natlinkbreak{\NAT@aysep\NAT@spacechar}{\@citeb\@extra@b@citeb}%
     \NAT@date}}
  {\@citea\NAT@nmfmt{\NAT@nm}%
   \NAT@aysep\NAT@spacechar\NAT@hyper@{\NAT@date}}{}{}

\patchcmd{\NAT@citex}
  {\@citea\NAT@hyper@{%
     \NAT@nmfmt{\NAT@nm}%
     \hyper@natlinkbreak{\NAT@spacechar\NAT@@open\if*#1*\else#1\NAT@spacechar\fi}%
       {\@citeb\@extra@b@citeb}%
     \NAT@date}}
  {\@citea\NAT@nmfmt{\NAT@nm}%
   \NAT@spacechar\NAT@@open\if*#1*\else#1\NAT@spacechar\fi\NAT@hyper@{\NAT@date}}
  {}{}

\makeatother

\begin{document}

\title{Type III Solar Radio Burst Source Region Splitting Due to a Quasi-Separatrix Layer}
        
\author{Patrick I. McCauley\tsp{1}, Iver H. Cairns\tsp{1}, John Morgan\tsp{2}, Sarah E. Gibson\tsp{3}, James C. Harding\tsp{1}, Colin Lonsdale\tsp{4}, and Divya Oberoi\tsp{5}}
\affil{\tsp{1}School of Physics, University of Sydney, Sydney, NSW 2006, Australia \\ 
	\tsp{2}International Centre for Radio Astronomy Research, Curtin University, Perth, WA 6845, Australia \\
	\tsp{3}National Center for Atmospheric Research (NCAR), Boulder, CO 80301, USA \\
	\tsp{4}MIT Haystack Observatory, Westford, MA 01886-1299, USA \\
	\tsp{5}National Centre for Radio Astrophysics, Tata Institute for Fundamental Research, Ganeshkhind, Pune 411007, India}
\email{patrick.mccauley@sydney.edu.au}
	
\shorttitle{MWA Type III Source Splitting}
\shortauthors{McCauley et al.}

\begin{abstract}

We present low-frequency (80--240 MHz) radio imaging of type III solar radio bursts observed  
by the \textit{Murchison Widefield Array} (MWA) on 2015/09/21.
The source region for each burst splits from one dominant component at higher frequencies into two increasingly-separated 
components at lower frequencies. 
For channels below $\sim$132 MHz, the two components repetitively diverge 
at high speeds (0.1--0.4 c) along directions tangent to the limb, with each episode 
lasting just $\sim$2 s.
We argue that both effects result from the strong magnetic field connectivity gradient that the 
burst-driving electron beams move into. 
Persistence mapping of extreme ultraviolet (EUV) jets observed by 
the \textit{Solar Dynamics Observatory} reveals quasi-separatrix layers (QSLs) associated with coronal null points, 
including separatrix dome, spine, and curtain structures. 
Electrons are accelerated at the flare site toward an open QSL, where the beams follow diverging field lines to produce the source splitting, with larger separations at larger heights (lower frequencies). 
The splitting motion within individual frequency bands is interpreted as a projected time-of-flight effect, whereby electrons 
traveling along the outer field lines take slightly longer to excite emission at adjacent positions. 
Given this interpretation, we estimate an average beam speed of 0.2 c. 
We also qualitatively describe the quiescent corona, noting in particular that 
a disk-center coronal hole transitions from being dark at higher 
frequencies to bright at lower frequencies, turning over around 120 MHz. 
These observations are compared to synthetic images based on the 
Magnetohydrodynamic Algorithm outside a Sphere (MAS) model, which we use to flux-calibrate the burst data. 

\end{abstract}



\section{Introduction}
\label{intro}

Type III solar radio bursts are among the principal signatures of magnetic reconnection, the process thought to  
underlie solar flares. 
Their high brightness temperatures demand a coherent, nonthermal emission 
mechanism that is generally attributed to plasma emission stimulated by semi-relativistic electron beams.
Electrons accelerated at the reconnection site generate Langmuir waves (plasma oscillations) in the ambient plasma through the 
bump-on-tail beam instability. 
Those Langmuir waves then shed a small fraction of their energy in radio emission near the 
fundamental plasma frequency ($f_{p}$) or its second harmonic. 
This theory was proposed by \citet{Ginzburg58} and has since been developed by many authors (see reviews by \citealt{Robinson00,Melrose09}). 

Radio bursts are classified by their frequency drift rates,  
and type IIIs are so named because they drift faster than types I and II \citep{Wild50}. 
A recent review of type III literature is provided by \citet{Reid14}. 
Starting frequencies are typically in the 100s of MHz, and because 
the emission frequency is proportional to the 
square of the ambient electron density ($f_{p}~\propto{}~\sqrt{n_{e}}$), 
standard type III radiation drifts to lower frequencies as the accelerated electrons stream outward. 
\textit{Coronal} type III bursts refer to those that drift down to tens of MHz or higher. 
Beams that 
escape along open field lines may continue to stimulate Langmuir waves in the solar wind plasma, 
producing \textit{interplanetary} type III bursts that may reach 20 kHz and below around 1 AU and beyond. 
We will focus on coronal bursts for which some fraction of the electrons do escape to produce 
an interplanetary type III.  

X-ray flares and type III bursts have been linked by many studies. 
Various correlation rates have been found, with a general trend toward increased association with better instrumentation.  
Powerful flares ($\ge{}$C5 on the GOES scale) almost always generate coherent radio emission, generally meaning a  
type III burst or groups thereof \citep{Benz05, Benz07}. 
Weaker flares may or may not have associated type IIIs depending on the magnetic field configuration \citep{Reid17}, 
and type IIIs may be observed with no GOES-class event if, for instance,  
the local X-ray production does not sufficiently enhance the 
global background \citep{Alissandrakis15}. 
Flares that produce X-ray or extreme ultraviolet (EUV) jets are frequently    
associated with type III emission \citep{Aurass94,Kundu95,Raulin96,Trottet03,Chen13b,Innes16,Mulay16,Hong17,Cairns17}. 
Such jets are collimated thermal plasma ejections that immediately follow,  
are aligned with, and are possibly heated by 
the particle acceleration responsible for radio bursts \citep{Saint-Hilaire09,Chen13}. 
We will exploit the alignment between EUV jets and type III 
electron beams to develop an understanding of radio source region behavior  
that, to our knowledge, has not been previously reported. 

This is the first type III imaging study to use the full 128-tile 
\textit{Murchison Widefield Array} (MWA; \citealt{Lonsdale09,Tingay13}), which 
follows from type III imaging presented by \citet{Cairns17} using the 32-tile prototype array. 
The MWA's primary 
science themes are outlined by \citet{Bowman13}, 
and potential solar science is further highlighted by \citet{Tingay13b}. 
The first solar images using the prototype array and later the full array are detailed by \citet{Oberoi11} 
and \citet{Oberoi14}, respectively.
\citet{Suresh16} present a statistical study of single-baseline dynamic spectra, 
which exhibit the lowest-intensity solar radio bursts ever reported. 
We present the first time series imaging. 

Along with the \textit{Low Frequency Array} (LOFAR; \citealt{van13,Morosan14}), the MWA 
represents a new generation of low frequency interferometers capable of solar imaging.
Previous imaging observations at the low end of our frequency range were made by the 
decommissioned \textit{Culgoora} \citep{Sheridan72,Sheridan83} 
and \textit{Clark Lake} \citep{Kundu83} radioheliographs, along with the 
still-operational \textit{Gauribidanur Radioheliograph} \citep{Ramesh98, Ramesh05}. 
The high end of the MWA's frequency range overlaps with 
the \textit{Nan\c{c}ay Radioheliograph} (NRH; \citealt{Kerdraon97}), which has 
facilitated a number of type III studies referenced here.
 
This paper is structured as follows. 
\S\ref{obs} describes our observations and data reduction procedures. 
Our analyses and results are detailed in \S\ref{analysis}. 
\S\ref{forward} considers the quiescent corona outside burst periods, which we compare 
to synthetic images used to flux calibrate the burst data in \S\ref{flux}.
\S\ref{kinematics} characterizes the type III source region structure and motion, 
and the local magnetic field configuration is inferred using EUV observations in \S\ref{persistence}. 
In \S\ref{discussion}, our results are combined to produce an interpretation of the  
radio source region behavior. 
\S\ref{conclusion} provides concluding remarks. 


  \begin{figure}
 \epsscale{1.1}
 \plotone{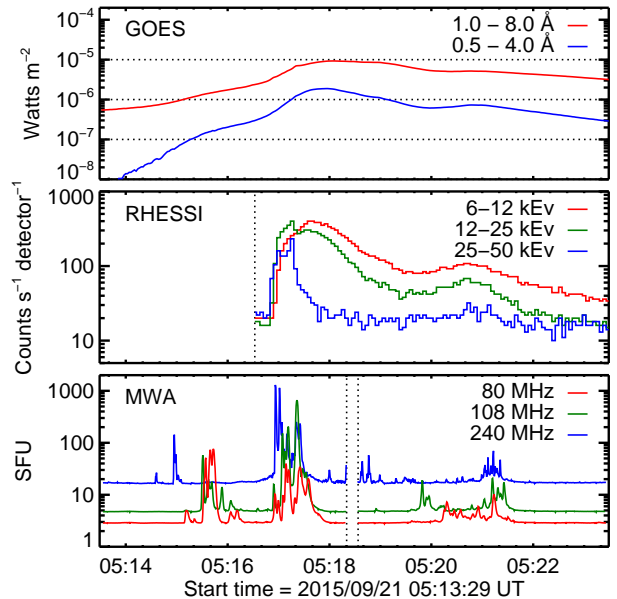}
 \caption{
\textit{Top}: GOES soft X-ray light curves, showing the C8.8 flare that peaked at 05:18 UT. Dotted lines 
from bottom to top indicate the B, C, and M-class thresholds. 
\textit{Middle:} RHESSI count rates from 6--50 kEv. The dotted line indicates the end of RHESSI's night (Earth-eclipse) period. 
\textit{Bottom:} MWA light curves at 80, 108, and 240 MHz. Dotted lines indicate the transition between continuous 
observing periods. 
}
 \label{fig:lightcurves}
 \end{figure}


  \begin{figure}
 \epsscale{1.1}
 \plotone{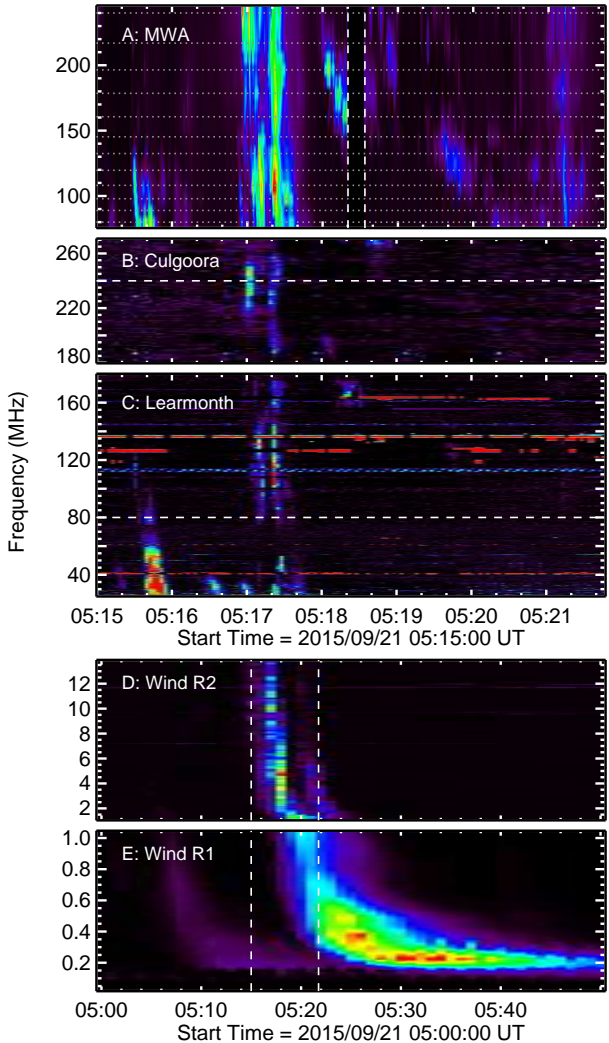}
 \caption{
\edit{\textit{A}}: MWA dynamic spectrum (DS) produced from total image intensities 
and interpolated to a spectral resolution equal to the minimum separation between observing bandwidths (see \S\ref{mwa}.) 
Dashed vertical lines indicate the transition between continuous observing periods, and dotted horizontal lines mark 
the 12, 2.56 MHz-wide frequency channels.
\edit{\textit{B--C}}: Culgoora and Learmonth DS. 
Dashed lines indicate the MWA frequency coverage bounds (80--240 MHz).
\edit{\textit{D--E}}: \textit{Wind}/WAVES RAD\edit{2 and RAD1} DS. Note that the time axis is expanded to show the 
low-frequency tail. The dashed \edit{lines indicate the period covered by panels A--C}.
All DS are log-scaled and then background-subtracted.
A corresponding movie is available in the 
\href{http://www.physics.usyd.edu.au/~pmcc8541/mwa/20150921/}{online material}.
}
 \label{fig:spectra}
 \end{figure}


\section{Observations}
\label{obs}

We focus on a brief series of type III bursts associated with a C8.8 flare that peaked at 05:18 UT on 
2015/09/21. 
The flare occurred in Active Region 12420\footnote{\href{https://www.solarmonitor.org/index.php?date=20150921\&region=12420}{AR 12420 summary @ solarmonitor.org}} 
on the east limb. 
This investigation began by associating MWA observing periods that utilize the 
mode described in \S\ref{mwa} with isolated type III bursts logged   
in the National Oceanic and Atmospheric Administration (NOAA) solar event 
reports\footnote{\href{http://www.swpc.noaa.gov/products/solar-and-geophysical-event-reports}{NOAA event reports @ swpc.noaa.gov}}. A small sample of bursts detected from 80 to 240 MHz were selected, 
and we chose this event for a case study because of the unusual source structure and motion. 
A survey of other type III bursts is ongoing. 

Figure~\ref{fig:lightcurves} shows the soft X-ray (SXR) light curves from the 
\textit{Geostationary Operational Environmental Satellite} (GOES\footnote{\href{http://www.swpc.noaa.gov/products/goes-x-ray-flux}{GOES X-ray flux @ swpc.noaa.gov}})
for our MWA observation period, along 
with those from the \textit{Reuven Ramaty High-Energy Solar Spectroscopic Imager} (RHESSI; \citealt{Lin02}). 
The corresponding MWA light curves, as derived in \S\ref{mwa} and \S\ref{forward}, show that the radio bursts occur primarily around the hard X-ray (HXR, 25--50 keV) peak and just before the SXR peak, with some minor radio bursts scattered throughout the SXR rise and decay phases. 
HXR and type III emissions are known to be approximately coincident in time \citep{Arzner05} and  
are generally attributed to oppositely-directed particle acceleration, with HXR 
production resulting from heating by the sunward component. 
The same process may underlie both, however small differences in the timing, along with  
large differences in the requisite electron populations, suggest there may be multiple related acceleration processes (e.g. \citealt{Brown77,Krucker07,White11,Cairns17}). 
In contrast, SXR emission is associated with thermal plasma below the reconnection site, generally peaking somewhat later with a 
more gradual profile as in Figure~\ref{fig:lightcurves}.

Our initial radio burst detections relied on observations from the \textit{Learmonth} and \textit{Culgoora} solar radio spectrographs. 
Part of the global \textit{Radio Solar Telescope Network}\footnote{\href{https://www.ngdc.noaa.gov/stp/solar/solarradio.html}{RSTN data @ ngdc.noaa.gov}} (RSTN; \citealt{Guidice81}), the \textit{Learmonth} spectrograph covers 25 to 180 MHz 
in two 401-channel bands that run from 25--75 and 75--180 MHz. Additional technical details are 
provided by \citet{Kennewell03}. 
The \textit{Culgoora} spectrograph\footnote{\href{http://www.sws.bom.gov.au/World_Data_Centre/2/5}{Culgoora data @ sws.bom.gov.au}} \citep{Prestage94} 
has broader frequency coverage (18--1800 MHz) over four 501-channel bands.
Only the 180--570 MHz band is relevant here, and we show just a portion of it because the \textit{Learmonth} spectrograph 
is more sensitive where they overlap. 
Both instruments perform frequency sweeps every 3 s. 
Dynamic spectra are plotted in Figure~\ref{fig:spectra}, each being log-scaled and   
background-subtracted by 5-min boxcar averages. 

Figure~\ref{fig:spectra} also includes \edit{dynamic spectra} from the 
\textit{Radio and Plasma Wave Investigation} (WAVES; \citealt{Bougeret95}) 
on the \textit{Wind} spacecraft. These data demonstrate an 
interplanetary component to the coronal type III bursts, which requires there be connectivity to open field lines 
along which electrons escaped the corona. This will be important to our interpretation of the 
magnetic field configuration in \S\ref{discussion}. 


  \begin{figure*}
 \epsscale{1.1}
 \plotone{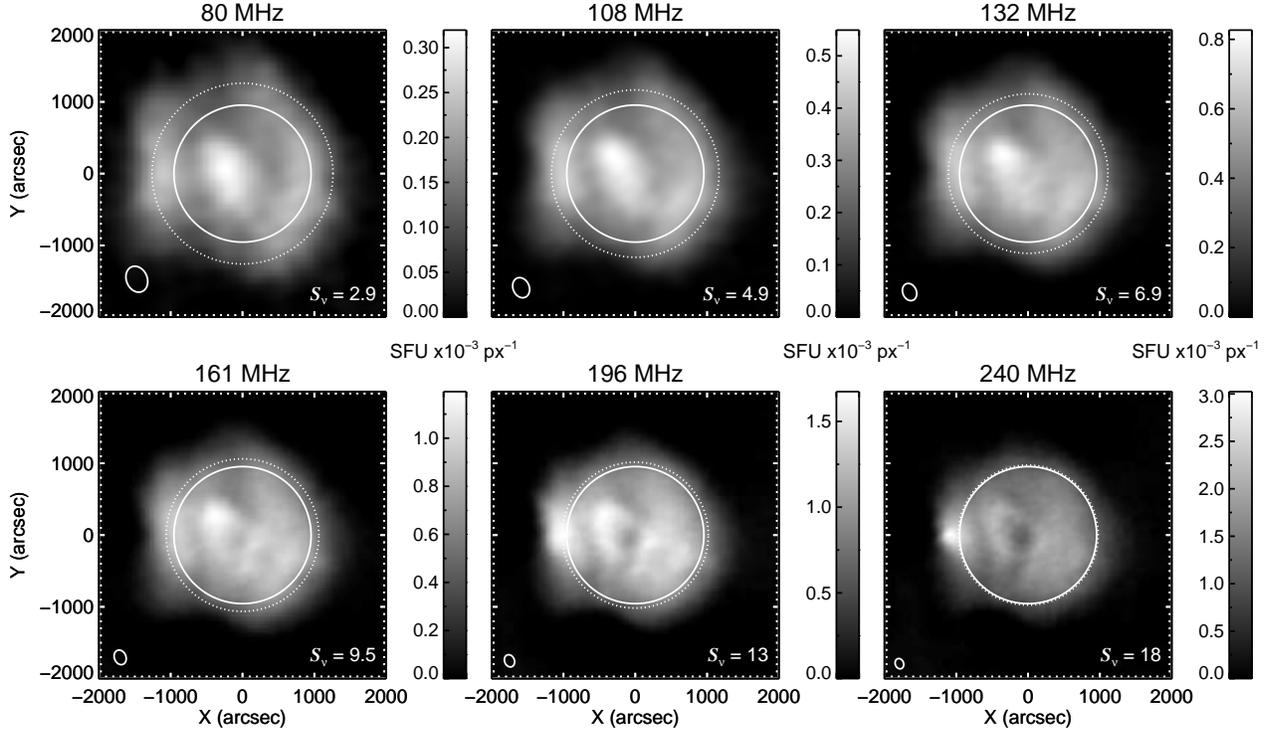}
 \caption{
MWA Stokes I images for 6 of the 12 frequency bands during a quiescent period at 2015/09/21 05:13:33.20 UT.
The solid inner circles denote the optical disk, and the dotted outer circles denote the 
Newkirk-model \citep{Newkirk61} limb for a given frequency. Ellipses in the bottom-left corners 
represent the synthesized beams. Values in the bottom-right corners are full-Sun integrated flux densities ($S_{\nu}$) in SFU, 
and the color bars represent the flux density enclosed by each \edit{20$\arcsec$} pixel in SFU$\times{}10^{-3}$ (see \S\ref{flux} for details). 
A movie showing the full time series for all 12 bands 
is available in the 
\href{http://www.physics.usyd.edu.au/~pmcc8541/mwa/20150921/}{online material}. 
 }
 \label{fig:preburst}
 \end{figure*}


  \begin{figure*}
 \epsscale{1.1}
 \plotone{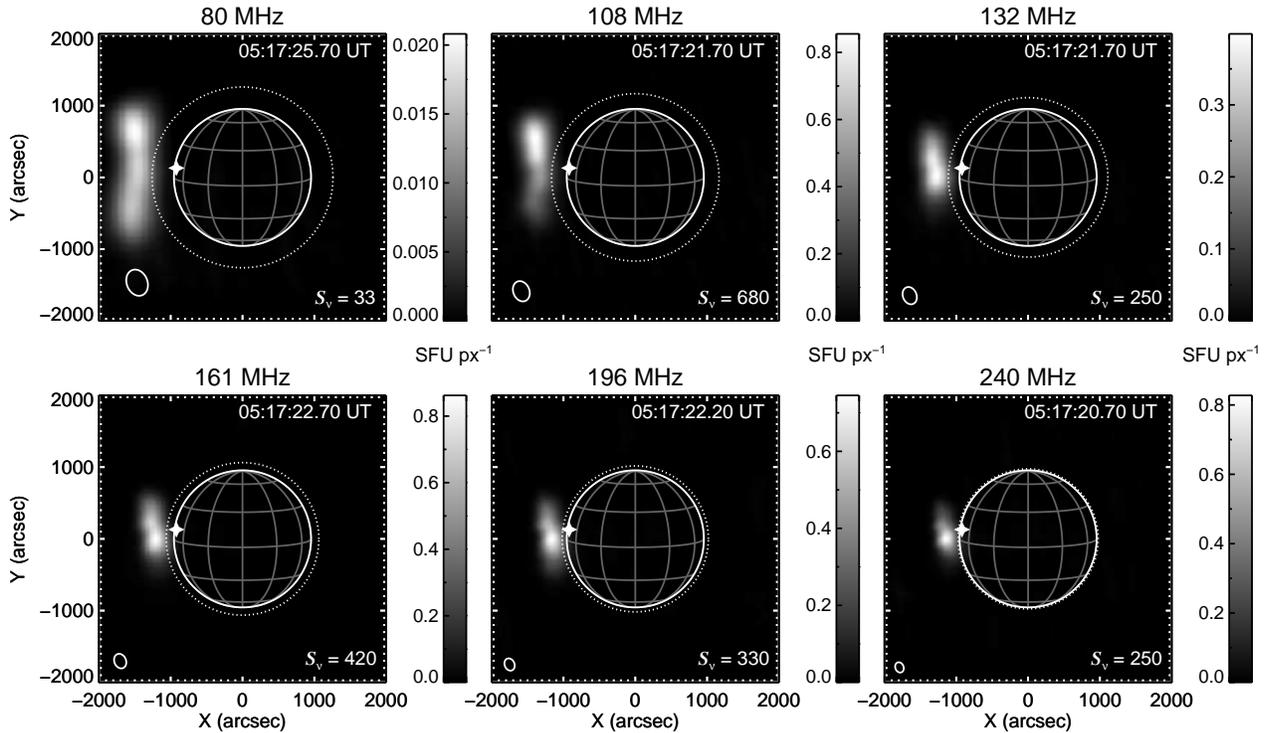}
 \caption{
Same as Fig.~\ref{fig:preburst} but for the \edit{frequency-specific peak intensity times associated with 
the event from 05:17:20 to 05:17:25 UT, which may comprise multiple overlapping bursts (see \S\ref{kinematics} \& \S\ref{discussion}}). 
Color bar units \edit{are} in SFU\edit{ px\tsp{-1}, and} stars mark the X-ray flare site.
 }
 \label{fig:burst}
 \end{figure*}


\subsection{Murchison Widefield Array (MWA)}
\label{mwa}

The MWA is a low-frequency radio interferometer in Western Australia that consists of 
128 aperture arrays (``tiles"), each comprised of 16 dual-polarization dipole antennas \citep{Tingay13}.
It has an instantaneous bandwidth of 30.72 MHz that can be spread flexibly from 80 to 300 MHz. 
Our data employ a ``picket fence" observing mode, whereby twelve 2.56 MHz bands are 
distributed between 80 and 240 MHz with gaps of 9--23 MHz between them. 
This configuration is chosen to maximize spectral coverage while 
avoiding radio frequency interference (RFI).
Data are recorded with a time resolution of 0.5 s and a spectral resolution of 40 kHz, which we 
average across the 2.56 MHz bandwidths to produce images centered at 
80, 89, 98, 108, 120, 132, 145, 161, 179, 196, 217, and 240 MHz.
Figures \ref{fig:preburst} and \ref{fig:burst} 
show images at six frequencies during quiescent and burst phases, respectively, and 
a movie showing all twelve bands over the full time series is available in the 
\href{http://www.physics.usyd.edu.au/~pmcc8541/mwa/20150921/}{online material}\footnote{Movie links: \url{http://www.physics.usyd.edu.au/~pmcc8541/mwa/20150921/}}. 

Visibilities were produced using the standard MWA correlator \citep{Ord15} and \texttt{cotter} \citep{Offringa15}. 
For our calibrator observations, this included 8-s time averaging and RFI flagging 
using the \texttt{aoflagger} algorithm \citep{Offringa12}. 
RFI flagging was disabled for the solar observations, as it tends to flag out burst data. 
Calibration solutions for the complex antenna gains were obtained 
with standard techniques \citep{Hurley14} using observations 
of a bright and well-modelled calibrator source (Centaurus A) 
made $\sim$2 hours after the solar observations. 
To improve the calibration solutions, the calibrator was imaged and ten loops  
of self-calibration were performed in the manner described by \citet{Hurley17}. 

This last step is typically performed on science target images, but we  
apply it instead to the calibrator for two reasons. 
First, we find that day-time observations generally produce inferior calibration solutions 
compared to analogous night-time data. 
We attribute this to contamination of the calibrator field 
by sidelobe emission from the Sun, but ionospheric and temperature effects 
may also be important. 
Second, the \texttt{clean} algorithm 
essential to the self-calibration process works best when the field is dominated by compact, 
point-like sources, which is not the case for the Sun.
The same steps performed on our solar images tended to degrade the overall quality of the calibration 
solutions and bias the flux distribution of the final images. 
However, we find that it is best to self-calibrate on the field source to obtain 
quality polarimetry because transferring calibration solutions from a lower-elevation 
pointing typically produces overwhelming Stokes I leakage into the other Stokes portraits. 
For this reason, we do not include polarimetry here. 
Progress has been made on producing reliable polarimetric images of the Sun with the MWA, as well as 
improving the dynamic range, but that is beyond the scope of this paper.

Once calibrated, imaging for each 0.5 s integration is accomplished using \texttt{WSClean} \citep{Offringa14} with the 
default settings except where noted below. 
Frequencies are averaged over each 2.56 MHz bandwidth, excluding certain fine channels 
impacted by instrumental artefacts. 
To emphasize spatial resolution, we use the Briggs -2 weighting scheme \citep{Briggs95}. 
Cleaning is performed with $\sim$10 pixels across the synthesized beam, 
yielding 16--36$\arcsec$ px\tsp{-1} from 240--80 MHz. 
We use a stopping threshold of 0.01, which is roughly the average 
RMS noise level in arbitrary units obtained for quiescent images cleaned with no threshold. 
Major clean cycles are used with a gain of 0.85 (\texttt{-mgain 0.85}), and peak finding uses the 
quadrature sum of the instrumental polarizations (\texttt{-joinpolarizations}). 
Finally, Stokes I images are produced using the primary 
beam model described by \citet{Sutinjo15}. 

To compare MWA data with other solar imaging observations, we introduce the \texttt{mwa\_prep} 
routine, now available in the 
SolarSoftWare libraries for IDL 
(SSW\footnote{SSW: \url{https://www.lmsal.com/solarsoft/}}, \citealt{Freeland98}). 
\texttt{WSClean} and the alternative MWA imaging tools produce FITS images using the   
\textsc{sin}-projected celestial coordinates standard in radio astronomy. 
Solar imaging data typically use ``helioprojective-cartesian" coordinates, which is a \textsc{tan} projection 
aligned to the solar rotation axis with its origin at Sun-center \citep{Thompson06}. 
To convert between the two coordinate systems, \texttt{mwa\_prep} rotates the image about 
Sun-center by the solar P angle, interpolates onto a slightly different grid to account for the 
difference between the \textsc{sin} and \textsc{tan} projections, and scales 
the images to a uniform spatial scale (20$\arcsec$ px\tsp{-1}). 
By default, the final images are cropped to 6$\times$6 $\rm{R}_{\odot}$, yielding 289$\times$289 pixels. 
FITS headers are updated accordingly, after which the various SSW mapping tools can 
be used to easily overplot data from different instruments.

We will consider quiescent radio structures in \S\ref{forward} against corresponding 
model images that are used for flux calibration in \S\ref{flux}. 
Burst structure and dynamics are discussed 
in \S\ref{kinematics}.  


  \begin{figure*}[h]
 \epsscale{1.1}
 \plotone{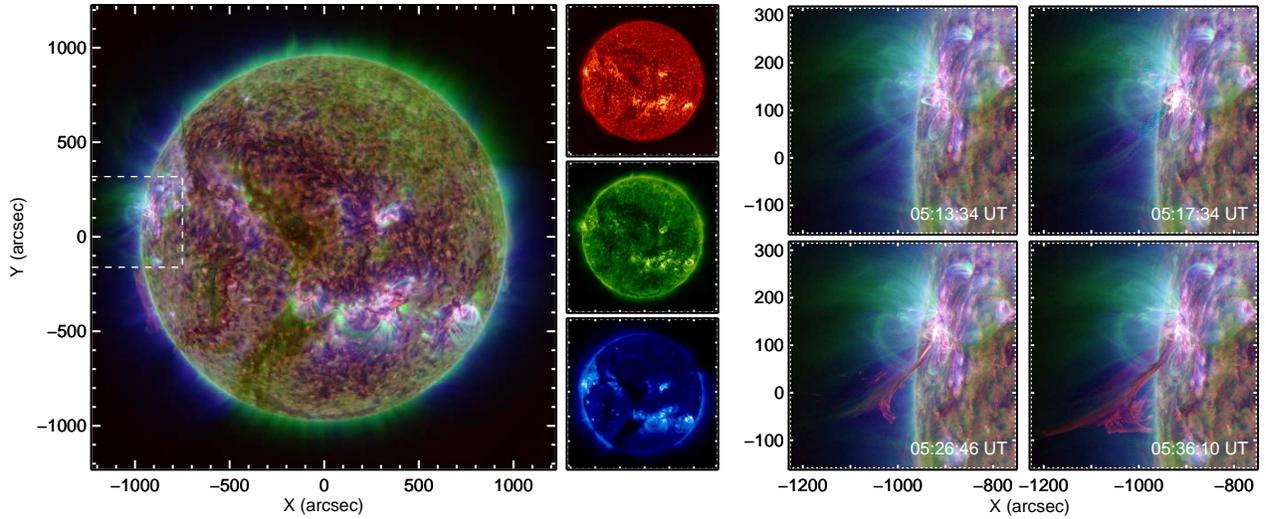}
 \caption{
An overview of the event seen by SDO/AIA using RGB composites of the 
304, 171, and 211 \AA{} channels. 
The top panels on the right half show nearly the same times as 
Figs.~\ref{fig:preburst} (left) and~\ref{fig:burst} (right), with the rightmost panel corresponding to 
just before the SXR peak. 
The bottom-right panels show snapshots of the EUV jets that follow 
the radio bursts. 
 }
 \label{fig:aia}
 \end{figure*}


\subsection{Solar Dynamics Observatory (SDO)}
\label{sdo}

The \textit{Solar Dynamics Observatory} (SDO, \citealt{Pesnell12}) is a satellite with three instrument suites, 
of which we use the \textit{Atmospheric Imaging Assembly} (AIA; \citealt{Lemen12}). 
We also indirectly use photospheric magnetic field observations from the \textit{Helioseismic and Magnetic Imager} (HMI; \citealt{Scherrer12}), 
which inform the synthetic images in \S\ref{forward}. 
The AIA is a full-Sun imager consisting of four telescopes that observe 
in seven narrowband EUV channels with a 0.6\arcsec{} px$^{-1}$ spatial resolution and 12 s cadence, 
along with three UV bands with a lower cadence. 

Calibrated (``level 1") data are obtained from the Virtual Solar Observatory 
(VSO\footnote{VSO: \url{http://sdac.virtualsolar.org/}}, \citealt{Hill09}). 
The SSW routine \texttt{aia\_deconvolve\_richardsonlucy} is used to deconvolve the images 
with filter-specific point spread functions, and \texttt{aia\_prep} is used to co-align and uniformly 
scale data from the different telescopes. 
Figure~\ref{fig:aia} presents an overview of our event using RGB composites of the 304, 171, and 211 \AA{} channels. 
These bands probe the chromosphere, upper transition region / low corona, and corona, respectively, 
with characteristic temperatures of .05 (He II), 0.63 (Fe IX), and 2 MK (Fe XIV). 

The AIA observations show a fairly compact flare that produces several distinct EUV jets beginning 
just before the soft X-ray peak at 05:18 UT. This includes higher-temperature material visible in up to the 
hottest band (94 \AA{}, 6.3 MK), along with cooler ejecta at chromospheric temperatures that appears 
in emission at 304 \AA{} and in absorption at other wavelengths. 
These outflows reveal a complex magnetic field configuration south of the flare site, which we 
will explore in \S\ref{persistence} and in \S\ref{discussion} with respect to the radio emission. 


\section{Analysis \& Results}
\label{analysis}


  \begin{figure*}
 \epsscale{1.1}
 \plotone{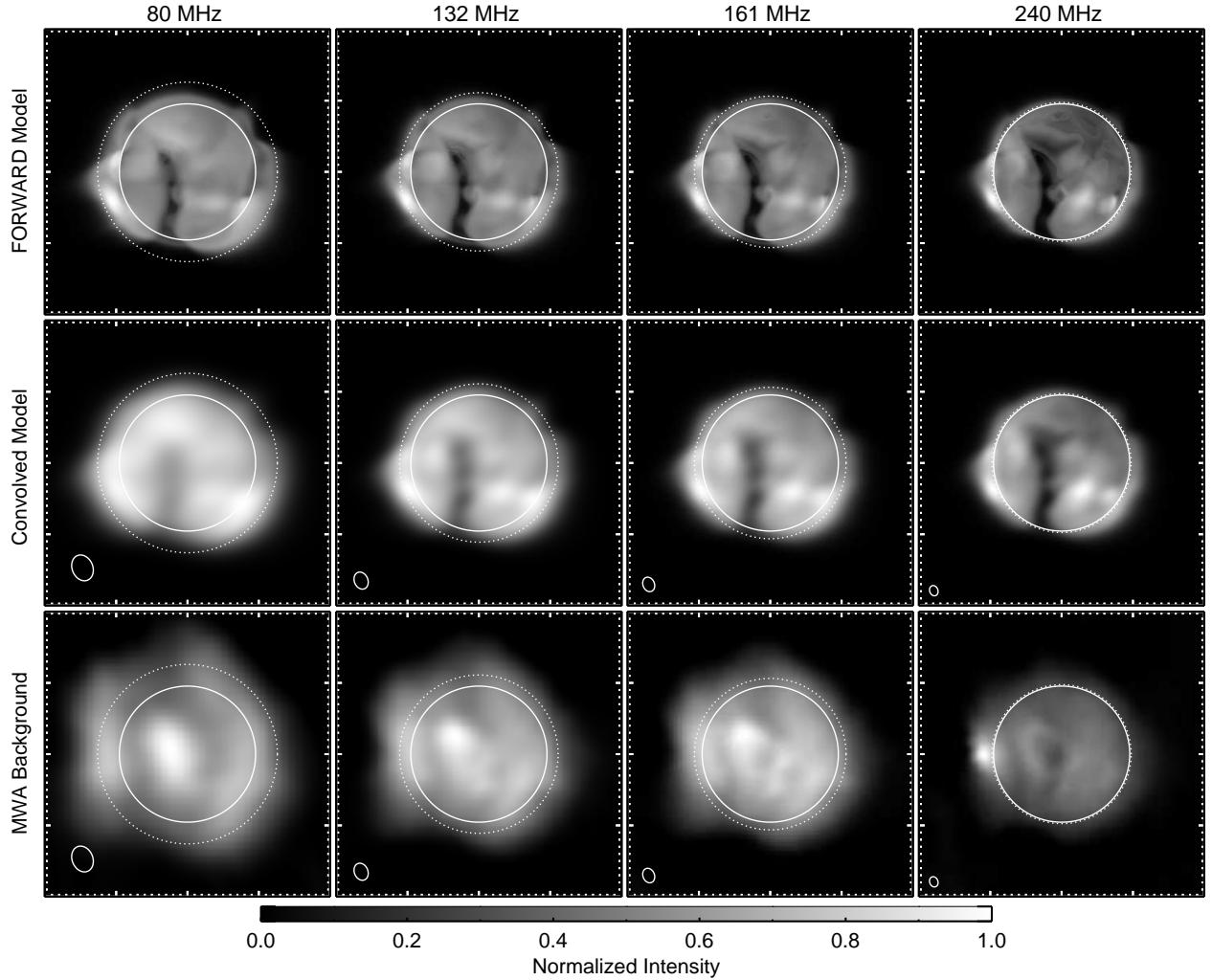}
 \caption{
\textit{Top}: Expected free-free and gyroresonance emission at four frequencies 
predicted by FORWARD based on the MAS thermodynamic MHD model.
\textit{Middle}: Model image convolved with the corresponding MWA beams. 
\textit{Bottom}: Median MWA emission outside burst periods over the first 4-min observation period, which is assumed to be 
the quiescent background for flux calibration. 
Plot axes and annotations are as in Fig.~\ref{fig:preburst}. 
An animation with all 12 channels is available in the \href{http://www.physics.usyd.edu.au/~pmcc8541/mwa/20150921/}{online material}. 
 }
 \label{fig:forward}
 \end{figure*}
 

  \begin{figure}
 \epsscale{1.1}
 \plotone{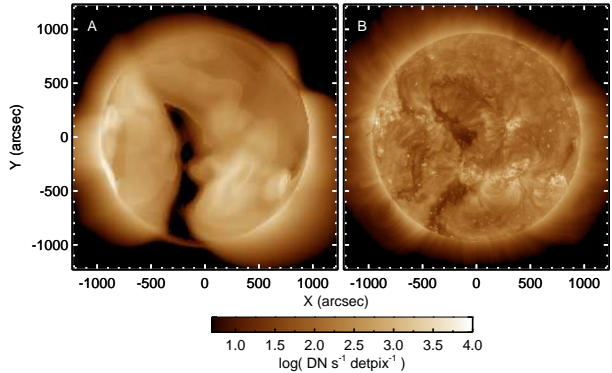}
 \caption{
193 \AA{} synthetic image (A) and SDO observation (B). The synthetic image applies the 
telescope response function so that both images are plotted on exactly the same scale 
in instrumental units (DN) per sec per detector pixel (detpix).  
 }
 \label{fig:forward_aia}
 \end{figure}
 

  \begin{figure*}
 \epsscale{1.1}
 \plotone{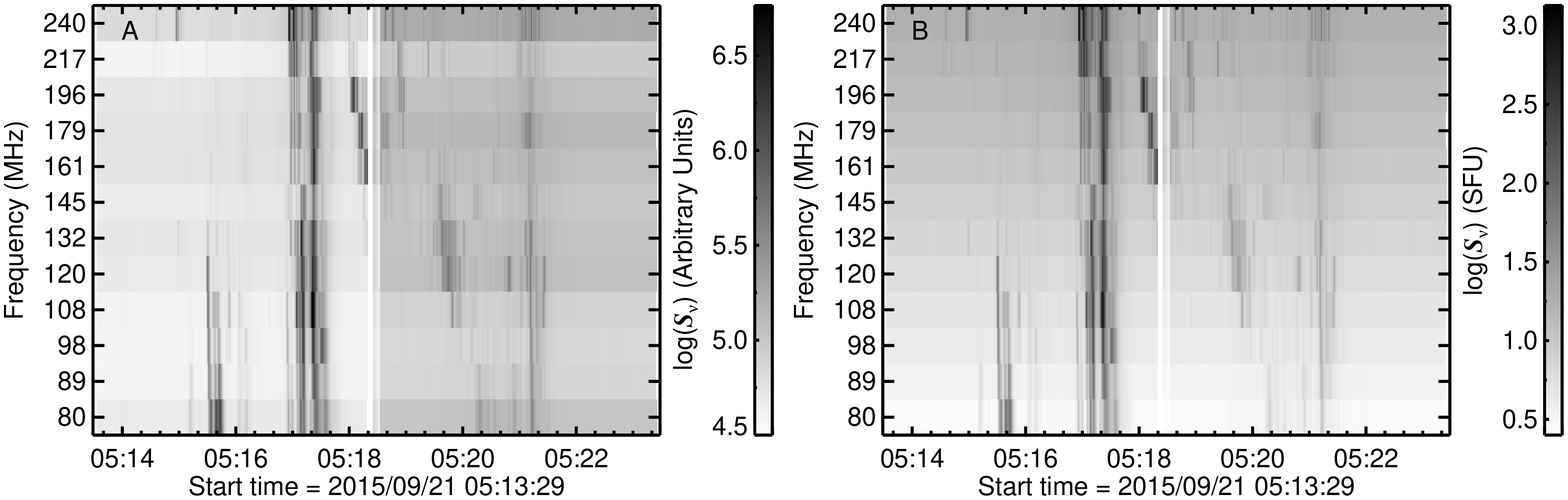}
 \caption{
Uncalibrated (A) and flux-calibrated (B) dynamic spectra generated from total image intensities. 
The Y axes intervals are not uniform; values refer to the 12 2.56-MHz-wide observing 
bandwidths separated by gaps of 9--23 MHz  (see \S\ref{mwa}). An interpolated dynamic spectrum with a uniform 
Y axis is shown in Fig~\ref{fig:spectra}. 
 }
 \label{fig:fluxcal}
 \end{figure*}


\subsection{Quiescent Structure \& Model Comparison}
\label{forward}

We examine model images of the coronal intensity at MWA frequencies to
qualitatively compare the expected and observed structures outside 
of burst periods. 
In the next subsection, we also use the predicted quiescent flux densities to obtain 
a rough flux calibration of our burst data. 
Synthetic Stokes I images are obtained using 
FORWARD\footnote{FORWARD: \url{https://www2.hao.ucar.edu/modeling/FORWARD-home}}, 
an SSW package 
that can generate a variety of coronal observables using different magnetic field and/or thermodynamic models. 
At radio wavelengths, FORWARD computes the expected contributions from thermal bremsstrahlung (free-free) 
and gyroresonance emission based on the modeled temperature, density, and magnetic field structure. 
Details on those calculations, along with the package's other capabilities, are 
given by \citet{Gibson16}.  

Our implementation uses the Magnetohydrodynamic Algorithm outside a Sphere 
(MAS\footnote{MAS: \url{http://www.predsci.com/hmi/data\_access.php}}; \citealt{Lionello09}) 
medium resolution (\texttt{hmi\_mast\_mas\_std\_0201}) model. 
The MAS model combines an MHD extrapolation of the coronal magnetic field (e.g. \citealt{Miki99}) based on 
photospheric magnetogram observations from the HMI with a heating model adapted from \citet{Schrijver04}. 
Comparisons between MAS-predicted images and data have been 
made a number of times for EUV and soft X-ray observations,  
with generally good agreement for large-scale structures (e.g. \citealt{Riley11,Reeves11,Downs12}). 
We make the first radio comparisons. 

The top row of Figure~\ref{fig:forward} shows synthetic images at four MWA frequencies.
Beam-convolved versions are shown in the middle row, but note that this does not 
account for errors introduced by the interferometric imaging process, such as effects 
related to deconvolving a mixture of compact and diffuse emission or to 
nonlinearities in the \texttt{clean} algorithm.  
MWA data are shown in the bottom row and 
reflect median pixel values over the first five-minute observation (05:13:33 to 05:18:20),   
excluding burst periods defined as when the total image intensities exceed 105\% of the first 0.5 s integrations 
for each channel. 
An animation with all 12 channels is available in the 
\href{http://www.physics.usyd.edu.au/~pmcc8541/mwa/20150921/}{online material}. 
For context, we also show a comparison of a 193 \AA{} SDO observation and prediction  
using the same model in Figure~\ref{fig:forward_aia}. 

The agreement between the observed and modelled radio images is best at our highest 
frequencies ($\gtrsim$ 179 MHz), where the correspondence is similar to that of the EUV case. 
For both, the model reproduces structures associated with coronal holes near the central meridian and 
the large active region complexes in the southwest. 
The large-scale structure associated with the southern polar coronal hole 
is also well-modelled for the radio case.
A similar structure is predicted for the EUV but is disrupted by the 
observed polar plumes in the manner described by \citet{Riley11}. 
The modelled images also under-predict emission from EUV coronal holes, 
which may be due to contributions from low-temperature ($<$ 500,000 K) material 
ignored by the emissivity calculations.
Other contributing factors might be inaccuracies in the heating model, evolution 
of the magnetic boundary from that used for the simulation, or 193 \AA{} emission 
from non-dominant ions formed at low temperatures. 

A number of discrepancies between the model and MWA observations are also apparent, 
particularly with decreasing frequency. 
With the exception of the bright region on the east limb at 240 MHz, which we will revisit in 
\S\ref{discussion}, we suspect these differences underscore the importance of 
propagation effects to the appearance of the corona at low frequencies. 
In particular, refraction (ducting) of radio waves as they encounter low-density regions,  
as well as scattering by density inhomogeneities, can profoundly alter the observed source 
structure (see reviews by \citealt{Lantos99,Shibasaki11}). 
Both effects can increase a source's spatial extent, decrease its brightness, and   
alter its apparent location (e.g. \citealt{Aubier71,Alissandrakis94,Bastian94,Thejappa08,Ingale15}).
We likely see the effects of scattering and/or refraction in the increased radial extent of 
the observed emission at all frequencies compared to the beam-convolved model images, though 
an enhanced density profile may also contribute. Likewise, these propagation effects 
may be responsible for dispersing the signatures of the southwestern active regions, which 
are prominent in the synthetic images but only barely discernible in our  
observations. 

Most conspicuously, the disk-center coronal hole gradually transitions from a dark feature at high frequencies 
to a bright one at low frequencies in the observations but not in the synthetic data. 
This could be due to the diminished spatial resolution at low frequencies, 
meaning the coronal hole signature is  
swamped by emission from the bright region to the northeast. 
However, that effect should serve only to reduce the coronal hole contrast,  
as it does for the beam-convolved synthetic images. 
Indeed, another set of observations of a different disk-center coronal hole also show this 
dark-to-bright transition from high to low frequencies with even less ambiguity. 
In both cases, the transition is gradual and turns over around 120 MHz. 
Above the $\sim$120 MHz transition we observe, coronal holes are consistently 
reported as intensity depressions (e.g. \citealt{Mercier12}), which is expected given their low densities. 
At longer wavelengths, coronal holes have sometimes been seen in emission \citep{Dulk74,Lantos87}, as 
in our lower frequency channels. 
Again, scattering \citep{Riddle74,Hoang77} and/or refraction \citep{Alissandrakis94} may be able to explain 
low-frequency enhancements in low-density regions, but a satisfactory explanation 
has not been achieved, in part because of limited data.
The MWA appears to be uniquely poised to address this topic given that the transition of certain 
coronal holes between being dark or bright features occurs within the instrument's frequency range, 
but an analysis of this is beyond the scope of this paper. 


\subsection{Flux Calibration}
\label{flux}

Absolute flux calibration is challenging for radio data because of instrumental 
uncertainties and effects related to interferometric data processing. 
Astrophysical studies typically use catalogs of known sources to set the flux scale, 
and many MWA projects now use results from the GaLactic and Extragalactic All-sky MWA Survey (GLEAM; \citealt{Hurley17}).
We cannot take this approach because calibrator sources are not distinguishable in close proximity 
to the Sun given the dynamic range of our data. 
Even calibrators at sufficiently large angular separations from the Sun to be imaged are 
likely to be contaminated by solar emission due to the MWA's wide field of view (see \S\ref{mwa}). 

To express our burst intensities in physical units, we 
take brightness temperature images from FORWARD and convert them to 
full-Sun integrated flux densities ($S_{\nu}$), which 
we then assume to be equal to the total flux density in the quiescent background images from Figure~\ref{fig:forward}. 
From this comparison, we obtain a simple multiplicative scaling factor to convert between the uncalibrated image intensities  
and solar flux units (SFU; 1 SFU = 10\tsp{4} Jy = 10\tsp{-22} W m\tsp{-2} Hz\tsp{-1}). 
This procedure is performed separately for both observing periods, and 
Figure~\ref{fig:fluxcal} illustrates the result by plotting an uncalibrated dynamic spectrum next to the calibrated version. 

In the calibrated spectrum, we see that the quiescent intensities are coherently ordered in the pattern expected for 
thermal emission, with flux density increasing with frequency. 
Importantly, the adjacent MWA observing periods are also  
set onto very similar flux scales. 
We find an overall peak flux density of 1300 SFU at 240 MHz. 
Relative to the background, however, the burst series is most intense around 108 MHz, peaking at 680 SFU around 
140$\times$ the background level (see the log-scaled and then background-subtracted dynamic spectrum in Figure~\ref{fig:spectra}). 
This makes our event of moderate intensity compared to those in the literature (e.g. \citealt{Saint-Hilaire13}). 

This technique provides a simple way to obtain reasonable flux densities for radio bursts in 
order to place them generally in context.
Given the differences between the observations and synthetic images, this method should 
not be applied if very accurate flux densities are important to the results, which is 
not the case here. 
It would also not be appropriate for analyzing quiet-Sun features, nor for cases where non-thermal 
emission from a particular active region dominates the Sun for the entire observation period.
However in this case, we see primarily thermal emission that we suspect is modulated 
by propagation effects not considered by FORWARD. 
These effects are not expected to dramatically affect the total intensity but may decrease it somewhat, which would 
cause our flux densities to be overestimated. 

A more sophisticated solar flux calibration method has recently been developed by 
\citet{Oberoi17}, who use a sky brightness model 
to subtract the flux densities of astronomical sources, leaving just that produced by the Sun. 
This method is applied to data from a single short baseline, yielding a total flux density 
that can be used to calibrate images with a scaling factor analogous to ours.  
This approach would be appropriate for quiet-Sun studies and preferable for burst studies that make 
significant use of the fluxes.
We note that our method yielded quiescent fluxes within a factor of 2 of those found by \citet{Oberoi17}
for a different day, after accounting for the different polarizations used. 
Future work will explicitly compare the two approaches. 


  \begin{figure}
 \epsscale{1.1}
 \plotone{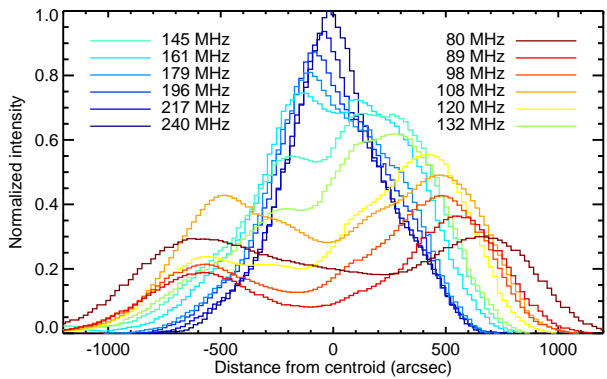}
 \caption{
Image slit intensities for each of the 12 MWA channels along the elongation axes of the individual burst source regions, 
illustrating the splitting of the source region from high to low frequencies.  
These data correspond to a period when the source regions are maximally extended at 05:17:26.6 UT. 
Each curve is normalized and multiplied by a scaling factor from 0.3--1.0 for clarity. 
 }
 \label{fig:elliptical_lightcurves}
 \end{figure}


  \begin{figure*}
 \epsscale{1.1}
 \plotone{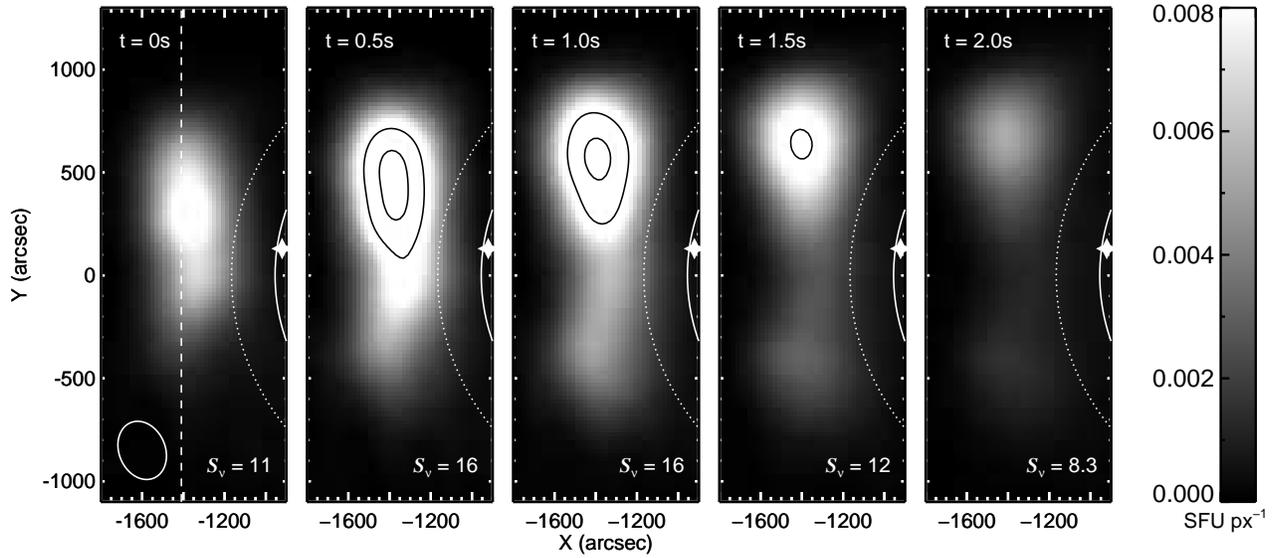}
 \caption{
Source splitting motion at 108 MHz, beginning at \edit{05:16:53.70} UT. The dashed line in the left 
panel denotes the slit used in Fig.~\ref{fig:kinematics}. 
The two solid black contours in the source region are 
at 0.010 and 0.015 SFU px\tsp{-1}. 
Additional annotations are as in Fig.~\ref{fig:burst}.
}
 \label{fig:splitting}
 \end{figure*}
 

  \begin{figure*}
 \epsscale{1.1}
 \plotone{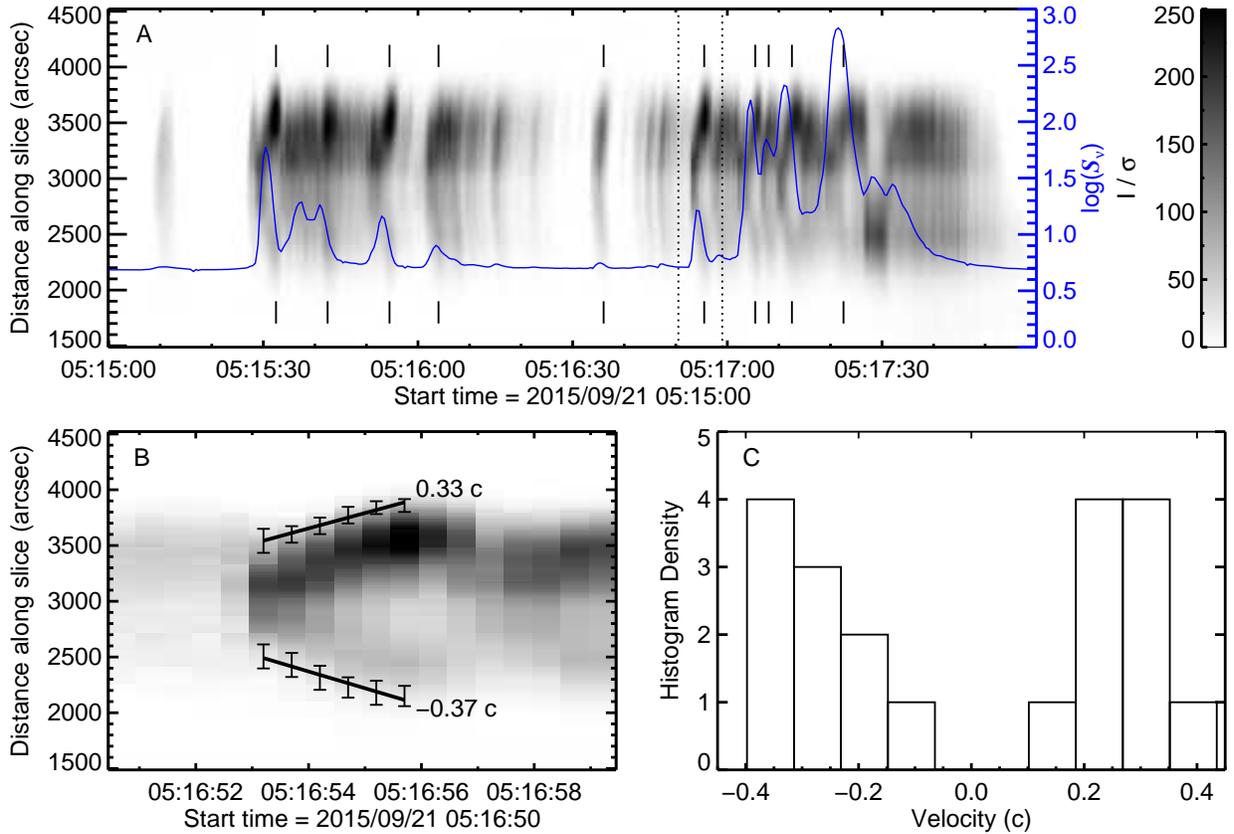}
 \caption{
An overview of the source splitting kinematics at 108 MHz. 
Panel A shows a distance-time plot using the slit shown in Fig.~\ref{fig:splitting} 
along with a light curve of the total flux density in blue.
Dotted vertical lines demarcate the zoomed-in section in Panel B\edit{, 
which corresponds to the images shown in Fig.~\ref{fig:splitting}.} 
Vertical ticks mark the 10 speed measurement periods whose results are collected 
in Panel C. Error bars in Panel B reflect the range of leading-edge estimates, 
obtained by thresholding the two components by 15--25\% of 
their maximum $I\cdot{}\sigma{}^{-1}$ values. 
 }
 \label{fig:kinematics}
 \end{figure*}


\subsection{Type III Source Structure \& Motion}
\label{kinematics}

The type III bursts begin around 05:15:30 UT during the early rise phase of the 
X-ray flare and continue at intervals through the decay phase. 
The two main bursts distinguishable in the \textit{Learmonth} and \textit{Culgoora} 
spectrographs are approximately 
coincident with the hard X-ray peak around 05:17 UT (Figure~\ref{fig:lightcurves}). 
The more sensitive and temporally-resolved MWA observations reveal these events to 
have a complicated dynamic spectrum structure that we interpret as the overlapping 
signatures of multiple electron injections in a brief period (Figure~\ref{fig:spectra}).

Throughout all of the bursts, a consistent pattern emerges in both the spatial structure of the source 
regions as a function of frequency and in their motions at particular frequencies. 
At higher frequencies, the type III source region is dominated by one spatial component 
with a much fainter component immediately to the north. 
Moving to lower frequencies and correspondingly larger heights, the two components 
separate along a direction tangent to the limb, reaching a peak-to-peak separation of 
1200$\arcsec$ (1.25 R$_{\odot}$) at 80 MHz. 
This structure is clear from the burst images in Figure~\ref{fig:burst} and is illustrated in 
further detail by Figure~\ref{fig:elliptical_lightcurves}. 

Figure~\ref{fig:elliptical_lightcurves} plots intensities extracted from image slits along the 
directions for which the emission is maximally extended. 
Slit orientations are determined by fitting ellipses to the overall  
source region in each channel after thresholding the images above 20\% of their peak intensities. 
Distances refer to that from the ellipse centers along their major axes, with values 
increasing from south to north. 
For clarity, the intensities are normalized and then multiplied by arbitrary scaling factors between 0.3 and 1.0 
from low to high frequencies.  
At least two Gaussian components are required to fit the curves at all frequencies, though 
the northern component is manifested only as a non-Gaussian shoulder on the dominant 
component at high frequencies. 
At some frequencies (e.g. 108 MHz), there are also additional weaker peaks between 
the two main components. 
Interpretation of the varying burst morphology as a function of frequency is given in \S\ref{discussion}. 

The type III source region components also 
spatially diverge as a function of time within single-channel observations below $\sim$132 MHz. 
At higher frequencies, for which there are one or two closely-spaced components, the source 
regions instead become increasingly elongated with time.
The direction of this motion is essentially the same as that of the frequency-dependent splitting, and 
the timescales for it are quite short, on the order of $\sim$2 s. 
This motion is repeated many times throughout the event, with each burst and corresponding 
``split" interpreted as a distinct particle acceleration episode. 
An example image set is shown in Figure~\ref{fig:splitting} for 108 MHz, 
the frequency that exhibits the highest intensities relative to the background.

To quantify this behavior, we employ distance-time 
maps to track movement along a particular slice through the images. The emission along the slit shown 
in the left panel of Figure~\ref{fig:splitting} is extracted from each observation and stacked against 
those from adjacent images, such that each vertical column of Figures~\ref{fig:kinematics}a and 
~\ref{fig:kinematics}b represents the slit intensity at a given time. Slopes in the ``slit image" 
correspond to plane-of-sky velocity components in the slit direction. 
Figure~\ref{fig:kinematics}a shows the result of this analysis for the bursts during the 
first MWA observation period, lasting nearly 3 minutes 
after 05:15 UT. Intensities have been divided by the time-dependent noise level, defined as the standard 
deviation of values within a 5-pixel-wide border around the edge of each image 
(equivalent in area to a 75$\times$75 px, or 25$\times$25 arcmin, box).
Because the noise level is roughly 
proportional to the total intensity, which varies by 2--3 orders of magnitude, this operation flattens the 
dynamic range of the distance-time map and provides for the uniform thresholding scheme described 
next. 

\edit{Throughout the series, the bursts peak in intensity at around the midpoint in the splitting 
motion, which is illustrated by the blue light curve in Figure~\ref{fig:kinematics}a. 
When the motion ends, 
the source regions gradually fade into the background with constant morphologies, 
or they are supplanted by those of a subsequent burst.
This decay phase manifests as the flat region in the distance-time profile in Figure~\ref{fig:kinematics}b. 
Note that the time period for Figure~\ref{fig:elliptical_lightcurves} is chosen so that each of the frequencies are in the 
declining phase, which is possible in that case because a subsequent burst does not follow for several 
seconds.}

The leading edges of the two source regions (north and south) are defined and tracked independently by 
thresholding the slit image 
above a percentage of the peak signal-to-noise ratio (SNR) for each component.
Measurements are made for each burst using 11 integer thresholds between 
between 15 and 25\% of the peak SNR.
\edit{This corresponds to values of 40--67 $\sigma$ for the northern component and 19--32 $\sigma$ for the southern.}
Error bars in Figure~\ref{fig:kinematics}b represent the resulting range of leading edge locations, and 
corresponding speed uncertainties are on the order of 15\%. 
An SNR percentage is used instead of a single set of values for both sources 
because it expands the range of reasonable thresholds, better representing the measurement 
uncertainties compared to a more restrictive range that would be appropriate for both sources. 

\edit{We also explored quantifying the same motion by instead tracking the centroid positions of the 
two source components. 
This approach was ultimately discarded because of difficulties in reliably  
separating the two main components across the full time series, particularly when the region is most 
compact at the beginning of each burst.
Our results may be hindered somewhat by scattering of the 
type described in \S\ref{forward}, which will be most pronounced near the source region perimeter. 
However, this would only affect the measured speeds if the scattering properties change significantly over 
the distance covered, and there appears to be little deviation of the leading edge slope from 
that of the overall source pattern in Figure~\ref{fig:kinematics}b.}

Vertical ticks in Figure~\ref{fig:kinematics}a mark the 10 bursts for which speed measurements were 
made at 108 MHz, and a histogram of the results is plotted in Figure~\ref{fig:kinematics}c. 
The time periods were chosen for particularly distinct source 
separation for which both components could be tracked. 
It is clear from Figure~\ref{fig:kinematics}a 
that the splitting motion occurs over a few additional periods for which measurements 
were precluded by confusion with adjacent events, faintness, or duration. 
We find speeds ranging between 0.11 and 0.40 c, averaging 0.26 c 
for the northern component and 0.28 c for the southern. The southern component is consistently 
faster for the 6 measurements before 05:16:55 UT and consistently slower after, but these differences 
are not statistically significant. 
These values cannot be straightforwardly interpreted as the exciter or 
electron beam speed \edit{(i.e. the average speed of accelerated electrons)} because that would require electrons traveling along flux tubes parallel  
to the limb in a manner inconsistent with the inferred magnetic field configuration (\S\ref{persistence}).
In \S\ref{discussion}, we will argue that this motion is a projected time-of-flight effect such that the splitting speeds 
here exceed the beam speed by a factor of $\lesssim$ 1.2. 

\begin{figure}
 \epsscale{1.1}
 \plotone{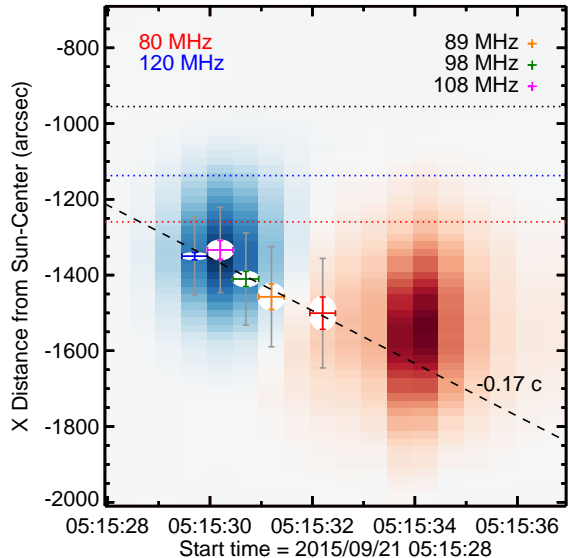}
 \caption{
Distance-time plot for burst emission from 05:15:28 to 05:15:37 UT. 
Red (80 MHz) and blue (120 MHz) images represent background-subtracted intensities averaged  
in the solar-Y direction, 
such that the slope reflects overall source motion in the solar-X direction. 
Crosshairs denote the burst onset times and centroid positions for each given frequency, where the 
onset is defined as exceeding 5$\times$ the background, 
Error bars correspond to the 0.5s time resolution (horizontal), the 3$\sigma{}$ variation in position 
over the burst period (vertical), and the minor synthesized beam axes (vertical, grey).  
Dotted horizontal lines represent the optical limb (black) and the Newkirk-model limbs at 
80 (red) and 120 (blue) MHz. 
}
 \label{fig:radial_extent}
 \end{figure}

\edit{The beam speed may be estimated more directly by examining} the burst location at different frequencies 
as a function of time. 
We do this in Figure~\ref{fig:radial_extent}, which shows a distance-time plot similar to Figure~\ref{fig:kinematics}. 
Instead of the emission along a particular slit, each column of Figure~\ref{fig:radial_extent} corresponds 
to the total image intensity binned down to a single row. 
Pixels with the same horizontal X 
coordinate are averaged, \edit{and} 
these \edit{Y-}averaged curves are stacked vertically against each other to show movement in the X direction. 
\edit{This is done so that the bidirectional vertical motion, 
which is primarily exhibited in single-channel observations (Figures~\ref{fig:splitting} \& \ref{fig:kinematics}), 
can be ignored to track the outward progression of the overall source region across frequency channels.}
Since our source regions are distributed on either side of the equator, this roughly corresponds to 
radial motion \edit{in the plane of the sky}. 

To \edit{quantify this motion}, we track the center position at the onset of the burst for 
each channel, which we define as 5$\times$ the background intensity.
\edit{We use the onset as opposed the times of peak intensity to avoid potential confusion 
between fundamental and second harmonic emission. Previous studies have shown from both 
observational \citep{Dulk84} and theoretical \citep{Robinson94} perspectives that emission at 
the fundamental plasma frequency arrives before associated harmonic emission, which may follow 
around the overall peak time after a frequency-dependent offset. 
Tracking the position at the onset of the burst thus ensures that we follow a coherent progression.
Note, however, that there is no standard in the literature. 
Estimates of type III beam speeds 
using the frequency drift rate technique, which will be discussed in \S\ref{discussion}, 
have used both onset and peak times (see review by \citealt{Reid14}).
}

Center positions are determined by fitting a Gaussian to the relevant time column. 
\edit{We track center positions here because the same difficulties described 
for Figure~\ref{fig:kinematics} do not exist in this case and also because it mitigates the potential 
influence of frequency-dependent scattering.
Scattering may still impact our result if the source locations are modulated significantly 
as a function of frequency, but we cannot readily test that possibility.}
We choose to examine the earliest burst period, occurring from 05:15:29--05:15:35 UT 
at frequencies below $\sim$132 MHz, because 
that event can be easily followed from high to low frequencies, whereas the more intense bursts  
later appear to \edit{comprise} several overlapping events. 
Fitting a line to the resulting spatiotemporal positions in Figure~\ref{fig:radial_extent}, we find a 
speed of 0.17 c. This result reflects the average outward motion of the entire source, 
which can be taken as a lower limit to the exciter speed. 

In comparison, the 108 MHz splitting speed for the same period averages to 0.28 c for both components, which  
as we will discuss in \S\ref{discussion}, \edit{exceeds the beam speed by a small factor based on the field geometry.} 
Thus we have a range of 0.17--0.28 c for the burst from 05:15:29--05:15:35 UT. 
Note that although the speeds from Figures~\ref{fig:kinematics} and~\ref{fig:radial_extent} 
are measured in orthogonal directions, we cannot combine 
them in a quadrature sum as though they were components of one velocity vector. 
As we will explain next, this is because we interpret 
the source behavior in terms of several adjacent electron beams, each with a slightly 
different trajectory than the next, as opposed to one coherent system. 
\edit{Also note that in all cases, we are estimating two-dimensional (plane-of-sky) velocity components 
of three-dimensional motion, which has a somewhat greater magnitude depending on the projection geometry.
Given this event's position on the limb and the direction of the EUV jets considered in the next section, we assume that 
the line-of-sight component is much smaller than its plane-of-sky counterpart.}


  \begin{figure*}
 \epsscale{1.1}
 \plotone{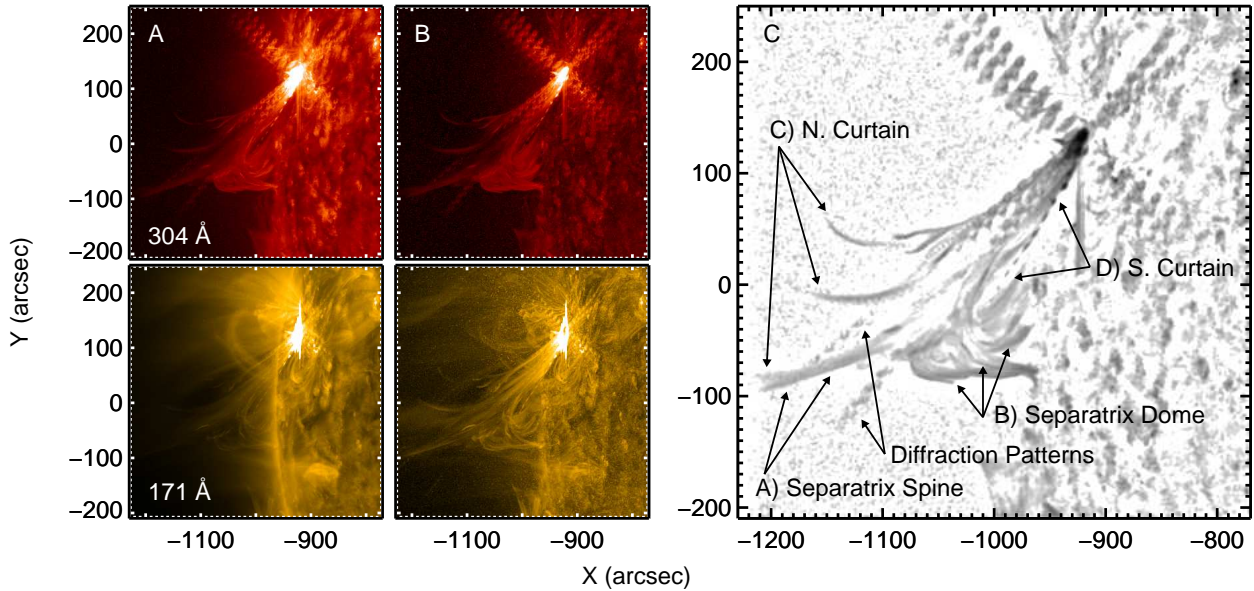}
 \caption{
A) Maximum value persistence maps for AIA 304 (top) and 171 \AA{} (bottom).
B) Column A subtracted by median backgrounds. 
C) Annotated 304-\AA{}, background-subtracted persistence map, further processed  
to accentuate features. See \S\ref{persistence} for processing details and 
Fig.~\ref{fig:cartoon} for a corresponding cartoon model. 
A corresponding movie is available in the \href{http://www.physics.usyd.edu.au/~pmcc8541/mwa/20150921/}{online material}.
 }
 \label{fig:pmap}
 \end{figure*}


\subsection{Magnetic Field Configuration}
\label{persistence}

Electron beams responsible for type III bursts propagate along magnetic field lines from 
the reconnection site, and therefore understanding the magnetic field configuration is critical to understanding 
the radio source region behavior and vice versa. 
Active region 12420, where the flare occurs, had just rotated into visibility on the east limb at the 
time of this event. 
EUV jets that immediately follow the radio bursts after the flare peak reveal a complex 
magnetic field configuration that connects AR 12420 to a small, diffuse dipole to the 
south near the equator.
The southern region was just behind the limb during the flare, and based on its evolution in HMI 
magnetograms over the following days, appears to have been a decaying active region 
near the end of its evolution. 

Unfortunately, this system is a poor candidate for local magnetic field modeling because of its 
partial visibility and position on the limb, where magnetogram observations are hampered by projection effects. 
The east limb position prevents us from using data from a few days prior, which is a possibility for 
west-limb events, and the decay of the southern dipole, along with the emergence of a neighboring region, 
dissuades us from attempting any dedicated modeling using 
data from subsequent days. 
Fortunately, the EUV jets trace out the field structure to an extent that we believe is sufficient to 
understand our observations. 
Previous studies have also demonstrated that type III electron beams are aligned with corresponding 
EUV and X-ray jets (e.g. \citealt{Chen13}), meaning that field lines traced out by the jets are 
preferentially those traversed by the accelerated electrons. 

We employ maximum-value persistence mapping to compile the separate EUV jet paths into one image. 
This style of persistence map refers simply to plotting the largest value a given pixel achieves over 
some period \citep{Thompson16}. 
Our maps cover from 05:18 to 05:39 UT, which corresponds to 
when the EUV jets begin around the peak flare time until they reach their full spatial extent visible to AIA around 
20 minutes later.
To further enhance the contrast, we subtract the persistence maps by a median-value background 
over the same period (i.e. $I_{\rm max} - I_{\rm med}$). 
Figures~\ref{fig:pmap}a and ~\ref{fig:pmap}b show maximum-value and background-subtracted 
persistence maps for both the 304 and 171 \AA{} channels, which are most sensitive to the jet material. 
Figure~\ref{fig:pmap}c shows a version of the 304 \AA{} map that has been Fourier filtered 
to suppress noise using a Hann window and then sharpened using an unsharp mask to accentuate 
the structure. 

The EUV jets trace out a toplogy, not apparent just 
prior to the flare, where the field connectivity changes rapidly. 
Such regions are generally known as \textit{quasi-separatrix layers} (QSLs; \citealt{Priest95,Demoulin96}), 
which are 3D generalizations of 2D separatrices that separate magnetic field connectivity domains. 
The key distinction is that the field linkage across a QSL is not discontinuous 
as in a true separatrix but instead changes drastically over a relatively small spatial scale, 
which can be quantified by the \textit{squashing factor} $Q$ \citep{Titov07}. 
QSLs are important generally because they are preferred sites for the development of current 
sheets and ultimately magnetic reconnection \citep{Aulanier05}. 
They are an essential part of 3D generalizations of the standard flare model \citep{Janvier13}, 
and modeling their evolution can reproduce a number of observed flare features (e.g. \citealt{Savcheva15, Savcheva16, Janvier16}). 
Here, we are less concerned with the dynamics of the flare site itself and focus instead on the 
neighboring region revealed by the EUV jets, which exhibits a topology associated 
with coronal null points. 
 
We first note that our observed structure is similar in several ways to that 
modeled by \citet{Masson12} and observed by \citet{Masson14}. 
The essential components are firstly the closed fan surface, or \textit{separatrix dome}, 
and its single \textit{spine} field line that is rooted in the photosphere and crosses the dome through 
the null point \citep{Lau90,Pontin13}. 
Open and closed flux domains are bounded above and below a separatrix dome, which can form 
when a dipole emerges into a preexisting open field region (e.g. \citealt{Torok09}). 
Above the dome and diverging around the null point is a vertical fan surface, or \textit{separatrix curtain}, 
comprised of field lines extending higher into the corona, with those closest to the separatrix spine likely being 
open to interplanetary space. 
Potential field source surface 
(PFSS; \citealt{Schrijver03})
extrapolations (not shown) do predict open field in this region but do not reproduce other topological features, which 
is to be expected given the modeling challenges described above. 
Some openness to interplanetary space must also have been present to facilitate the corresponding 
interplanetary burst observed by \textit{Wind} and shown in Figure~\ref{fig:spectra}. 

The separatrix dome, spine, and part of the curtain are clearly delineated by the EUV jets 
and are labeled in Figure~\ref{fig:pmap}c.
Note that some of the features, namely the closed field line associated with the southern 
portion of the separatrix curtain, are somewhat difficult to follow in Figure~\ref{fig:pmap}c but 
can be clearly distinguished in the corresponding movie available in the 
\href{http://www.physics.usyd.edu.au/~pmcc8541/mwa/20150921/}{online material}. 
In the following section, we will discuss how both types of source splitting described in \S\ref{kinematics} are 
facilitated by this topology. 


  \begin{figure}
 \epsscale{1.1}
 \plotone{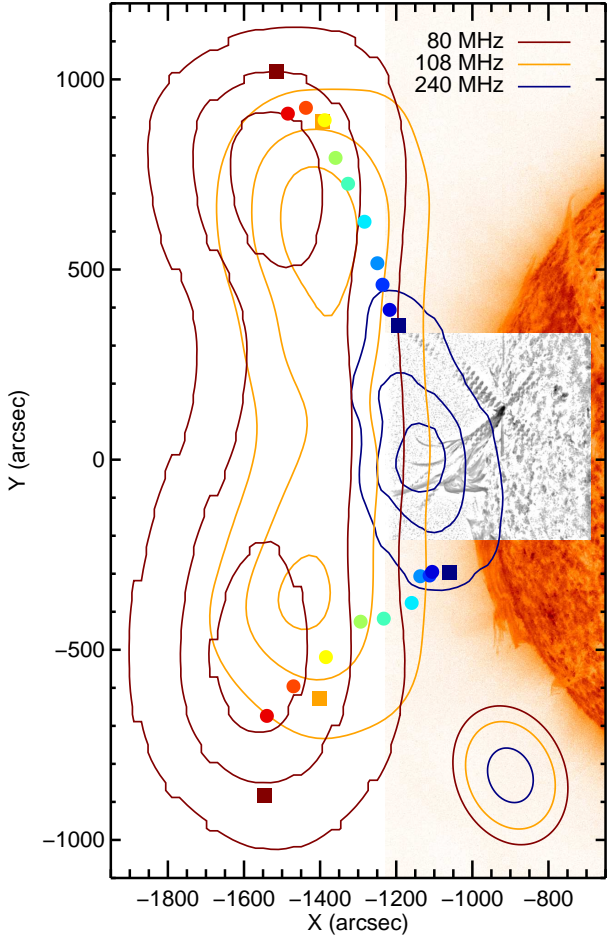}
 \caption{
MWA type III burst contours overlaid on a 304 \AA{} SDO image.
The greyscale inset is the persistence map from Fig.~\ref{fig:pmap}c. 
Pairs of colored dots represent the angular extent of the MWA source region in all 12 channels, 
with the squares from left to right corresponding to the \edit{reddish brown} (80 MHz), \edit{orange} (108 MHz), 
and \edit{dark} blue (240 MHz) contours, respectively. 
\edit{Contour levels are at 20, 50, and 80\% of the peak intensity.}
The MWA data are from a period when the source regions are maximally extended around 05:17:26.6 UT, and 
the SDO image combines data from the EUV jet period that follows (see \S\ref{persistence}).  
 }
 \label{fig:pmap_overlay}
 \end{figure}


  \begin{figure}
 \epsscale{1.1}
 \plotone{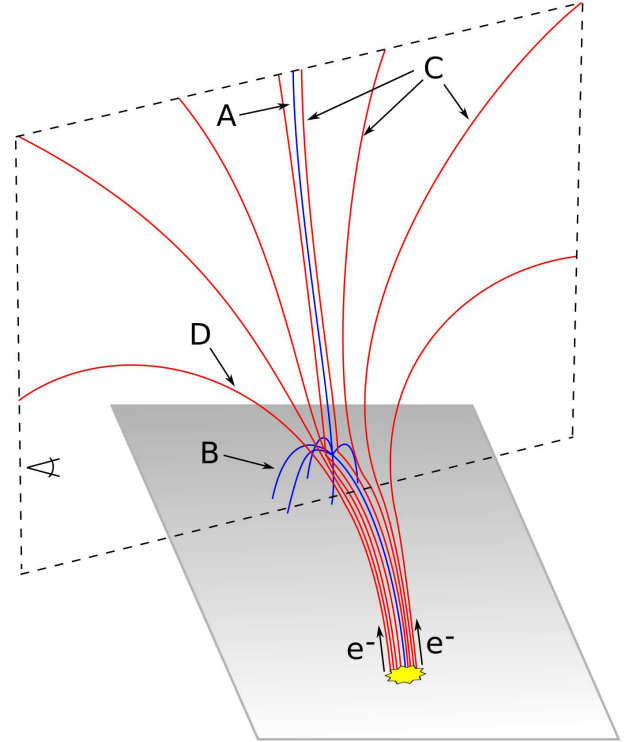}
 \caption{
Cartoon interpretation of the magnetic field configuration inferred from the EUV jet morphology and 
radio source regions (Fig.~\ref{fig:pmap_overlay}). 
The yellow region denotes the flare site, which is connected to a neighboring region with open and closed 
QSLs. 
Red field lines form a separatrix curtain, with the field closest to the center being open to interplanetary space. 
The blue field lines represent the closed separatrix dome, with a single spine field line that crosses the dome 
through a magnetic null point. 
Electrons travel along the diverging field lines of the separatrix curtain to produce the radio source structure and motion. 
Capital letters correspond to features apparent in the EUV observations (Fig.~\ref{fig:pmap}).  
 }
 \label{fig:cartoon}
 \end{figure}


  \begin{figure}
 \epsscale{0.9}
 \plotone{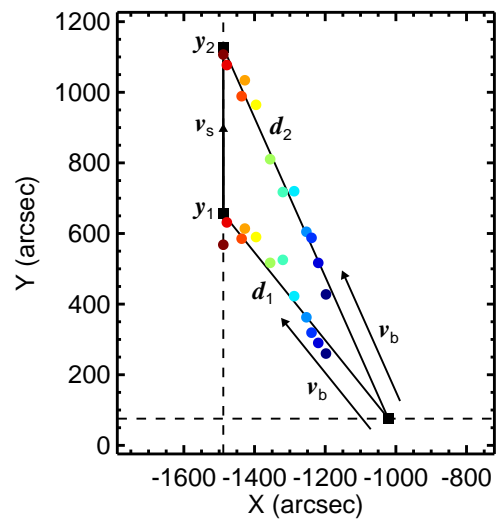}
 \caption{
Model schematic for the source splitting motion (Eqn.~\ref{eqn:split}).
Pairs of colored dots represent the average minimum and maximum vertical 
extents during each splitting episode; colors indicate frequency as in Figs.~\ref{fig:elliptical_lightcurves} \& \ref{fig:pmap_overlay}.
The flux tubes along which the type III beams travel are approximated by the solid fit lines,  
which intersect near the observed null point (Fig.~\ref{fig:pmap}). 
Electrons take slightly longer to reach $y_2$ compared to $y_1$, which produces the apparent 
vertical motion with velocity $v_{\rm s}$. 
In reality, there would be a number of adjacent curved flux tubes between and below the two lines with 
nearby, but not identical, origins. 
 }
 \label{fig:split_speed}
 \end{figure}


\section{Discussion}
\label{discussion}

When we overplot contours of the type III burst emission on the persistence map of 
the EUV jets (Figure~\ref{fig:pmap_overlay}), 
we see that the 240 MHz emission 
is concentrated just above the separatrix dome. 
As we described in \S\ref{kinematics}, the burst emission splits with decreasing frequency 
(increasing height) into two increasingly-separated components. 
Figure~\ref{fig:pmap_overlay} shows that the two components are distributed on 
either side of the separatrix spine. This implies a two-sided separatrix curtain with 
open field lines on either side of the spine, of which only the northern 
set are readily apparent in the EUV images. 
Given the position of the southern radio source and the closed field line that appears to form 
part of the southern curtain (D) in Figure~\ref{fig:pmap}, the southern half of the separatrix 
curtain seems to be oriented largely along the line of sight, which may explain why it is difficult 
to discern from the EUV jet structure. 
This two-sided separatrix curtain differs from the one-sided structure of \citet{Masson12,Masson14}, but a 
number of other studies consider somewhat similar topologies \citep{Maclean09,van12,Titov12,Craig14,Pontin15}. 

In Figure~\ref{fig:cartoon}, we sketch a 3D field configuration based on the aforementioned modeling 
studies that fits the EUV structure and extrapolates from there to satisfy the connectivity required by the 
radio source distribution.  
This cartoon can parsimoniously explain both the spatial splitting of the source from high to low frequencies 
and the source motion observed for individual frequency channels.
Type III bursts emit at the local plasma frequency or its second harmonic ($f \approx f_p$ or $2f_p$), 
which is proportional to the square of the ambient electron density. 
Thus, emission at a particular frequency can be associated with a particular height corresponding 
to the requisite background density. 
In our interpretation, electrons travel simultaneously along each of the red field lines 
in Figure~\ref{fig:cartoon}. 
The electron beams diverge on either side of the separatrix curtain, such that the beams are nearest 
to each other at lower heights (higher frequencies) and furthest apart at larger heights (lower frequencies). 
This produces the spatial source splitting and the dramatic increase of the overall angular extent toward 
lower frequencies, which is illustrated by the pairs of colored dots in Figure~\ref{fig:pmap_overlay}. 
\edit{The dots correspond to vertices of ellipses fit the overall source regions thresholded above 20\% of 
their peak intensities in the same manner and for same time period used in \S\ref{kinematics} for Figure~\ref{fig:elliptical_lightcurves}.} 

The source motions illustrated by Figures~\ref{fig:splitting} and~\ref{fig:kinematics} can then 
be accounted for as a projected time-of-flight effect. 
Electrons moving along the increasingly curved outer field lines take slightly longer 
to reach the same height, producing emission at adjacent positions along the separatrix curtain at 
slightly later times for a given frequency. 
\edit{This assumes that adjacent field lines have roughly the same radial density gradient, which implies  
decreasing density gradients along the field lines themselves as path lengths to specific heights (densities) 
increase with distance from the separatrix spine.} 
Thus, the splitting speeds measured in \S\ref{kinematics} are not the exciter or electron beam 
speeds. They are instead somewhat faster, depending on the difference 
in travel time to a given height along adjacent flux tubes. 
Adopting the geometry in Figure~\ref{fig:split_speed}, the 
expression for this is:

\begin{equation}
\label{eqn:split}
v_{\rm s} = \frac{y_2-y_1}{d_2-d_1}v_{\rm b}~, 
\end{equation}

\noindent{}where $v_{\rm s}$ is the apparent source splitting speed, $v_{\rm b}$ is the electron beam speed, 
$y_{1,2}$ are solar Y coordinates, and $d_{1,2}$ are the distances traveled along the 
field lines to reach $y_{1,2}$.

To estimate these parameters, we determine the average minimum and maximum vertical 
extents of the source regions for each frequency by fitting ellipses to every burst image, 
\edit{as was done for a single time step to illustrate the source region extents in Figures~\ref{fig:elliptical_lightcurves} and \ref{fig:pmap_overlay}.}
The X coordinates of the northern vertices are averaged, and the Y coordinates 
one standard deviation above and below the mean are averaged separately to 
obtain the pairs of colored dots in Figure~\ref{fig:split_speed}.
\edit{We take this approach rather than tracking the northern component's centroid 
because, along with the associated difficulties described in \S\ref{kinematics}, 
it allows us to capture consistent information from the higher-frequency channels where there is only one component 
and also because it is 
similar to the leading edge method used to estimate $v_{\rm s}$ in Figure~\ref{fig:kinematics}.}
 
If we approximate the field lines as linear fits to these points, which intersect close 
to the observed null point (Figure~\ref{fig:pmap}), then the speed of the source 
motion is 1.16$\times$ the beam speed.
Taking each of the lower-frequency points individually, we find factors ranging from 1.14 at 120 MHz 
to 1.19 at 80 MHz. 
Slightly larger factors are found for lower frequencies because of the larger 
separations between $y_{1}$ and $y_{2}$ compared to the fit projection, 
which may be due to the field lines curving out with height.

\edit{As with the $v_{\rm s}$ estimates in \S\ref{kinematics}, scattering may impact these results 
if the effect changes significantly between the colored dots in Figure~\ref{fig:split_speed}. 
Lower frequencies also tend to be more strongly scattered, which may enlarge the source regions 
as a function of decreasing frequency beyond the effect of the magnetic field divergence. 
Accounting for scattering would therefore preferentially decrease the Y-axis positions of the lower-frequency points 
in Figure~\ref{fig:split_speed}, which 
would flatten the slopes of both lines and slightly decrease the ratio $v_{\rm s}$/$v_{\rm b}$.
Including this effect would require an understanding of the local density structure and is beyond 
our scope. 
Also note that the model defined by Equation~\ref{eqn:split} and Figure~\ref{fig:split_speed} is 
specific to this magnetic field configuration and projection geometry. 
While the same basic effect may be observed for other events, different expressions may be needed to 
relate the observed motion to the beam speed.}

Using the 1.16 factor, \edit{the average speed ($v_{\rm s}$) from Figure~\ref{fig:kinematics}} 
corresponds to an average \edit{plane-of-sky beam speed ($v_{\rm b}$)} of 0.2 c. 
This value is consistent with and provides independent confirmation of beam speeds 
estimated from frequency drift rates, which is possible if one assumes a density model. 
Modest fractions of light speed are typical in the corona (e.g. \citealt{Alvarez73,Aschwanden95,Melendez99,Kishore17}), but some 
studies have found values in excess of 0.5 c \citep{Poquerusse94,Carley16} and even superluminal velocities 
given the right projection geometry \citep{Klassen03}. 
We also note that similar observations could be used to \edit{independently} probe the coronal 
density structure \edit{and beam speed} because our imaging capability allows us to estimate \edit{$v_{\rm b}$} without assuming a density model 
using time- and frequency-varying source positions in the manner illustrated by Figure~\ref{fig:radial_extent}. 
This particular event is not ideal for that analysis because of the complicated 
source structure, but a followup study is planned for a small ensemble of events that exhibit simple 
source structures without the type of motion described here. 
\edit{A similar study was also recently performed at lower frequencies (larger heights) by \citet{Morosan14} using type III 
imaging from LOFAR. They found speeds ranging from 0.3--0.6 c and observed emission at significantly larger 
heights than would be expected from standard density models.}

A few other connections to the literature should be mentioned with respect to the observed 
radio structure and inferred field configuration. 
First, we see from Figure~\ref{fig:forward} and in the movie associated with Figures~\ref{fig:preburst} 
and~\ref{fig:burst} that the source region of the bursts at 240 MHz is consistently 
enhanced and exhibits low-level burst activity outside of the intense burst periods. 
Figure~\ref{fig:pmap_overlay} demonstrates that this emission is concentrated just above the 
separatrix dome and associated null point.
These structures are interface regions between closed and open magnetic flux, where 
interchange reconnection may be ongoing (e.g. \citealt{Masson12,Masson14}). 
Such regions have previously been associated with radio enhancements and 
noise storms \citep{Wen07,Del11,Regnier13}. 

A few \textit{Nan\c{c}ay Radioheliograph} (NRH) observations exhibit characteristics reminiscent of  
those described here. 
For instance, \citet{Paesold01} conclude that the spatial separation of temporally adjacent 
type III events predominantly resulted from different field line trajectories followed by the 
electron beams. 
\citet{Reid14b} show a number of elliptically extended type III source regions that are represented 
as enveloping the diverging paths of electrons accelerated from the same site. 
Our observations that overlap in frequency with the NRH range ($\geq$150 MHz) are similarly extended to a larger degree before 
separating into two primary components at lower frequencies. 
\citet{Carley16} describe a ``radio arc" in their lowest-frequency images  
that is strikingly similar to our observations (e.g. Figure~\ref{fig:pmap_overlay}) but is suggested instead 
to trace the boundary of an erupting coronal mass ejection. 

\edit{We also} note that the complicated structure exhibited by the MWA dynamic spectrum 
(Figures~\ref{fig:spectra} \&~\ref{fig:fluxcal}) may indicate the presence 
of other burst types. Classic type III emission drifts from high to low frequencies as 
electron beams propagate outward into interplanetary space. 
If confined to closed field lines, the same beams may produce type U or J bursts 
for which the frequency drift rate switches signs as electrons crest the closed loops and 
propagate back toward the Sun \citep{Maxwell58,Aurass97,Reid17b}. 
We see hints of this in our dynamic spectrum at $\sim$196 MHz around 05:17:40 UT (Figure~\ref{fig:spectra}), 
but it is difficult to interpret because of the MWA's sparse frequency coverage. 
Given that our interpretation of the magnetic field configuration (Figure~\ref{fig:cartoon}) includes 
closed field lines on either side of the separatrix curtain, such features in the dynamic spectrum 
would not be surprising. Our splitting motion could also be due partially to beams traveling largely 
tangent to the limb along such closed field lines, while adjacent beams make it to larger heights along 
field lines closer to the separatrix spine, but evidence for downward propagation is lacking in the images. 

\edit{Finally, the bursts in this series do not all exhibit 
the statistical tendency for increasing type III flux densities with decreasing frequency 
(e.g. \citealt{Weber78, Dulk01, Saint-Hilaire13}), which is 
clear for the main event shown in Figure~\ref{fig:burst} and others visible in the flux-calibrated dynamic 
spectrum (Figure~\ref{fig:fluxcal}b).
Individual type III bursts often deviate from this pattern, exhibiting enhancements at particular 
frequencies or breaks in the emission over a particular frequency range. 
This behavior may be attributed to, among other things, density turbulence along the beam path \citep{Li12, Loi14} and/or
variations in the ambient electron and ion temperatures \citep{Li11b,Li11}. 
Additionally, electrons streaming along closed field lines, as considered in the previous paragraph, 
may contribute to enhancements at particular frequencies.}


\section{Conclusion}
\label{conclusion}

We have presented the first time series imaging study of MWA solar data.
Our observations reveal complex type III burst source regions that exhibit previously 
unreported dynamics. 
We identify two types of source region splitting, one being a frequency-dependent 
structure and the other being source motion within individual frequency channels. 
For the former, the source regions splits from one dominant component at our highest frequency (240 MHz) 
into two increasingly separated sources with decreasing frequency down to 80 MHz. 
This corresponds to a straightforward 
splitting of the source region as a function of height, with larger separations at larger heights. 

With high time resolution imaging, we observe a splitting motion within the source regions at 
individual frequencies, 
particularly in the lower channels ($\lesssim$ 132 MHz), 
that is tangent to the limb in essentially the same direction as the source splitting 
from high to low frequencies. 
This motion is short-lived ($\sim$2 s), fast (0.1--0.4 c), and repetitive, occurring multiple times over a period 
of 7 min before, during, and after the X-ray flare peak. 
We interpret the repetitive nature as multiple electron beam injections
that produce distinct radio bursts with overlapping signatures in the dynamic spectrum, which 
is consistent with there being 
several distinct EUV jet episodes that immediately follow the radio bursts. 

The EUV jets, which are assumed to have very 
similar trajectories to the type III electron beams, trace out a region where the 
magnetic field connectivity rapidly diverges over a small spatial scale. 
These types of configurations are broadly referred to as QSLs, 
and we argue that this field structure facilitates the radio source region splitting. 
Several common topological features associated with coronal null points are identifiable in 
persistence maps of the EUV outflows, including a separatrix dome, 
spine, and curtain. 
Electrons are accelerated simultaneously along adjacent field lines that connect 
the flare site to an open QSL, where their paths 
diverge to produce the source region splitting. 
At 240 MHz, the burst emission is concentrated just above 
the separatrix dome, a region that is consistently 
enhanced outside of burst periods. Moving to larger heights (lower frequencies), the source regions  
split on either side of the separatrix spine. 
\edit{The diverging field thereby enlarges the source regions at lower frequencies, an effect that 
may compound with angular broadening by refraction and scattering in this and other events.}
The northern radio component is consistent with field lines apparent from the EUV observations, 
but the southern component implies a two-sided separatrix curtain 
that is not obvious from the EUV observations. Thus, the radio imaging provides additional 
constraints on the magnetic field connectivity. 

The magnetic field configuration also offers a straightforward explanation for the radio source 
motion via a projected time-of-flight effect, whereby electrons moving along slightly longer outer field lines take slightly 
longer to excite emission at adjacent positions of roughly the same radial height. 
Given this interpretation, the speed of the source region is a factor of $\lesssim$ 1.2$\times$ greater than the 
electron beam speed. We estimate an average beam speed of 0.2 c, which is an independent confirmation of 
speeds estimated from frequency drift rates. 
We note that the same characteristics are observed in another type III burst from the same  
region three hours earlier.
This implies that the field topology 
is stable at least on that timescale and strengthens our conclusion 
that the radio dynamics are caused by interaction with 
a preexisting magnetic field structure, as opposed to peculiarities of the flare process itself. 

Lastly, we motivate future studies of MWA solar observations. 
A survey of type III bursts is underway. 
From preliminary results, we note that the dual-component splitting behavior described here is uncommon. 
However, analogous source region motion in one direction is common and could be explained in the 
same manner if coupled with a consistent picture of the particular field configurations. 
Similar events that occur near disk center or on the opposite (west) limb could be combined with magnetic field modeling 
to develop a more detailed topological understanding. 
The coronal density structure can also be probed by examining events with less complicated source structures. 
Finally, we showed a coronal hole that gradually transitions from dark to bright from high to low frequencies, turning over 
around 120 MHz. 
This adds a transition point to the small body of literature reporting coronal holes in emission at low frequencies, an 
effect that is not well-explained and could be addressed with additional MWA observations. 


\acknowledgements

PIM thanks Natasha Hurley-Walker for instruction on MWA data processing, 
Mike Wheatland and Yuhong Fan for discussions related to the magnetic field configuration, 
Emil Lenc for discussions related to polarization, 
and the Australian Government for supporting this work through an Endeavour Postgraduate Scholarship. 
\edit{We thank the anonymous referee for their constructive comments.}
JM, CL, and DO acknowledge support from the 
Air Force Office of Space Research (AFOSR) via grant FA9550-14-1-0192, 
and SEG acknowledges support from AFOSR grant FA9550-15-1-0030.
This scientific work makes use of the Murchison Radio-astronomy Observatory (MRO), operated by 
the Commonwealth Scientific and Industrial Research Organisation (CSIRO). 
We acknowledge the Wajarri Yamatji people as the traditional owners of the Observatory site. 
Support for the operation of the MWA is provided by the Australian Government's  
National Collaborative Research Infrastructure Strategy (NCRIS), 
under a contract to Curtin University administered by Astronomy Australia Limited. 
We acknowledge the Pawsey Supercomputing Centre, which is supported by the 
Western Australian and Australian Governments.
The SDO is a National Aeronautics and Space Administration (NASA) satellite, and 
we acknowledge the AIA and HMI science teams for providing open 
access to data and software. 
NCAR is supported by the National Science Foundation (NSF). 
This research has made use of NASA's Astrophysics Data System (ADS).
\linebreak \\
{\it Facilities:} 
\facility{MWA}; 
\facility{SDO (AIA)}; 
\facility{WIND (WAVES)}; 
\facility{GOES}; 
\facility{RHESSI}
 

\bibliographystyle{yahapj}
\bibliography{20150921_type3_references}  

\end{document}